\documentclass[11pt]{iopart}
\usepackage{graphicx}
\usepackage{color} 
\usepackage{transparent}
\graphicspath{{figure/}}

\usepackage{iopams}  

\begin{document}
\article[Contrast and phase-shift of a trapped atom interferometer...]{}{Contrast and phase-shift of a trapped atom interferometer using a thermal ensemble with internal state labelling}

\author{M. Dupont-Nivet$^{1,2}$\footnote{Corresponding author: matthieu.dupontnivet@thalesgroup.com}, C. I. Westbrook$^{2}$ and S. Schwartz$^{1}$}

\address{${}^{1}$Thales Research and Technology France, Campus Polytechnique, 1 av. Augustin Fresnel, 91767 Palaiseau, France}
\address{${}^{2}$Laboratoire Charles Fabry de l'Institut d'Optique, Campus Polytechnique, 2 av. Augustin Fresnel, 91127 Palaiseau, France}

\begin{abstract}
We report a theoretical study of a double-well Ramsey interferometer using internal state labelling. We consider the use of a thermal ensemble of cold atoms rather than a Bose-Einstein condensate to minimize the effects of atomic interactions. To maintain a satisfactory level of coherence in this case, a high degree of symmetry is required between the two arms of the interferometer. Assuming that the splitting and recombination processes are adiabatic, we theoretically derive the phase-shift and the contrast of such an interferometer in the presence of gravity or an acceleration field. We also consider using a "shortcut to adiabaticity" protocol to speed up the splitting process and discuss how such a procedure affects the phase shift and contrast. We find that the two procedures lead to phase-shifts of the same form. 
\end{abstract}

\vspace{2pc}

\maketitle

\section{Introduction}

Inertial sensors based on interferometry \cite{Kasevich1991} with freely falling atoms have demonstrated excellent performance in the measurement of gravity \cite{Peters1999}, gravity gradients \cite{McGuirk2002} and rotations \cite{Gustavson2000}. Atom interferometry with trapped atoms is much less well developed although it offers some advantages: interrogation times are not limited be the atoms' flight from the interaction region and one can hope to reduce the overall size of the device using technologies such as atom chips \cite{Schumm2005,Bohi2009,Fortagh2007}. These advantages motivated our recent proposal for a trapped atom interferometer using thermal atoms \cite{Ammar2014}, a situation closely analogous to white light interferometry in optics \cite{Lefevre2014}. In it we discussed the importance of maintaining a high degree of symmetry in the two interferometer arms. 

In that design we discussed use of internal state labeling of non-condensed ultra-cold atoms \cite{Bohi2009}, essentially a Ramsey interferometer with an adiabatic spatial separation of the internal states. An adiabatic procedure however, has the disadvantage of severely limiting the speed of the splitting: the separation must be slow compared to the trap oscillation period. Here we will consider another approach inspired by recent work on "shortcuts to adiabaticity" (STA) \cite{Torrontegui2013,Zhang2016} which allows one to effect the separation more rapidly \cite{Torrontegui2013,Schaff2011,Torrontegui2011}. This technique is already use in some experiments to move the position \cite{Torrontegui2011} and change the frequencies \cite{Chen2010} of a trap filled with a thermal gas or a Bose-Einstein condensate \cite{Schaff2011b}. Although a STA protocol is rather complex, we find that the resulting phase shifts and interferometer contrast are of the same intuitive form as in the adiabatic case.

In this paper we consider a protocol similar to the one described in reference \cite{Bohi2009,Ammar2014}, namely a Ramsey interferometer with spatial separation of the internal states. Such a configuration has the advantage of providing an independent control on the two arms of the interferometer \cite{Ammar2014}, and allows the phase to be measured by atom counting rather than fringe fitting. We take into account the possible effect of gravity or acceleration, and describe the dynamics of the splitting and recombination process in two particular cases. In the first case, we assume that the splitting and recombination process is slow enough that adiabatic approximation holds \cite{Ammar2014}. In the second case, we assume purely harmonic trap and derive an optimal interferometric sequence based on the shortcut to adiabadicity (STA) technique \cite{Torrontegui2013,Schaff2011}. 

This paper is organized as follows: in section \ref{sec_protocol}, we describe the basic principles of the interferometer protocol we consider. In section \ref{sec_adiabatic}, we discuss the phase-shift and contrast in the case of adiabatic splitting and recombination. In section \ref{sec_beyond_adiabatic}, we then consider the whole interferometric sequence as a dynamical problem, and show, in the case of harmonic potentials, that shortcuts to adiabaticity \cite{Torrontegui2013,Schaff2011,Torrontegui2011,Chen2010} can be used to reduce the splitting and recombination time. We give an expression for the dynamical phase-shift of the interferometer, including the effects of the slitting and recombination ramps, the temperature and the asymmetry between the trapping potentials.

\section{Interferometer protocol}
\label{sec_protocol}

In this section, we briefly recall the interferometer protocol described  in reference \cite{Ammar2014}, and that we will consider in the rest of this paper. Consider an ensemble of atoms with two levels $\left|a\right>$ and $\left|b\right>$. A typical interferometric sequence starts with a $\pi/2$ pulse to put the atoms in a coherent superposition of $\left|a \right>$ and $\left|b \right>$ with equal weights. Then the two internal states are spatially separated (the splitting period), held apart (the interrogation period) and recombined (the merging period) using state-dependent potentials $V_i(\widehat{z},t)$ which are only seen by atoms in internal state $\left|i\right>$. We note $\widehat{z}$ the position operator, $t$ the time and $i={a,b}$. We suppose that the design of the interferometer \cite{Ammar2014} allows $V_a=V_b$ at the beginning and at the end of the sequence. Finally, another $\pi/2$ pulse closes the interferometer. Between the two $\pi/2$ pulses, the system can be described by the following Hamiltonian \cite{Ammar2014}:
\begin{equation} \label{hamiltonian}
\widehat{H}= \frac{\widehat{p}^2}{2m}  +  V_{a}(\widehat{z},t) \left|a\right>\left<a\right| 
+ \left[V_b(\widehat{z},t)+\hbar\omega_{ab} \right] \left|b\right>\left<b\right| \;,
\end{equation}
where $\widehat{p}$ is the impulsion operator and $\hbar\omega_{ab}$ is the energy difference between the two internal states at the beginning and at the end of the interferometric sequence. Before the first $\pi/2$ pulse (labelled by $t=0$), we assume that the state of the atomic cloud is the same as in \cite{Ammar2014} (i.e. in the internal state $\left|a\right>$, at thermal equilibrium with temperature $T$ in the trapping potential $V_a$). Thus we describe it by the same density matrix $\widehat{\rho} = \sum_n p_n \left|n_a(0)\right>\left|a\right>\left<a\right|\left<n_a(0)\right|$. Here $n_a$ labels the energies levels in the trap $V_a$, the $p_n=e^{-E_n^a/kT}/\sum_n e^{-E_n^a/kT}$ are the Boltzmann factors where $E_n^a$ are the eigen-energies of $V_a(\widehat{z},0)$ and $\left|n_a(t)\right>\left|a\right>$ are solutions of the Schr\"odinger equation with the Hamiltonian $\widehat{H} \left|a \right>\left<a\right|(t)$ and constitute an orthonormal basis (the same notation will be used for $\widehat{H} \left|b \right>\left<b\right|(t)$ later on in the paper). As in \cite{Ammar2014} we neglect the effect of collisions in the atomic cloud during the interferometric sequence (i.e. we don't have damping term in the Liouville equation for the evolution of the density operator), thus, due to the choice of the $\left|n_i(t)\right>\left|i\right>$, the $p_n$ stay constant during the interferometric sequence. The effect of a $\pi/2$ pulse  is modelled by:
\begin{eqnarray}
\left|a\right> \rightarrow \frac{1}{\sqrt{2}}\left( \left|a\right> - ie^{-i\phi}\left|b\right> \right) \;, \qquad
\left|b\right> \rightarrow \frac{1}{\sqrt{2}}\left( \left|b\right> - ie^{+i\phi}\left|a\right> \right) \;.
\end{eqnarray}
where we have neglected the finite duration of the pulse, $\phi$ is the phase of the electromagnetic field at the beginning of the pulse, and $\omega$ the frequency of the electromagnetic field. This model is valid in the case $|\delta/\Omega| \ll 1$, where $\delta=\omega-\omega_{ab}$ is the detuning from the atomic resonance, and $\Omega$ is the Rabi frequency.

Just after the second $\pi/2$ pulse (labelled by $t=t_f$, where $t_f$ is the time between the two pulses), and using the hypothesis $V_a(\widehat{z},0)=V_b(\widehat{z},0)$ and $V_a(\widehat{z},t_f)=V_b(\widehat{z},t_f)$, the density matrix reads:
\begin{eqnarray} \label{rhotf}
\widehat{\rho} (t_f)  =  \sum_n p_n \left|n_a(t_f)\right> && \left\{  p_n^a \left|a\right>\left<a\right|  +  p_n^b \left|b\right>\left<b\right| \right.\nonumber\\
  && +  \left. p_n^{ab}\left|a\right>\left<b\right| + p_n^{ba}\left|b\right>\left<a\right|  \right\} \left<n_a(t_f)\right| \;,
\end{eqnarray}
with $p_n^a = \left[1-\cos{\left( \delta\phi - (\Omega_n^b-\Omega_n^a)  \right)}\right]/2$ and $p_n^b = \left[1+\cos{\left(\delta\phi -(\Omega_n^b-\Omega_n^a) \right)}\right]/2$ and $\delta\phi=\omega t_f$. Where $\Omega_n^i$ includes the dynamic and geometrical phases accumulated by $\left|n_i(t)\right>\left|i\right>$ between the two $\pi/2$ pulses.
In the above expressions, $p_n^i$ is the population of $\left|n_i(t)\right>$ in internal state $\left|i\right>$, and $p_n^{ab}$ and $p_n^{ba}$ are the coherence terms between the two internal states in level $\left|n_a(t)\right>$ and $\left|n_b(t)\right>$. As in \cite{Ammar2014}, the physical quantity measured in this interferometer is the total population in each internal state. We choose to write the total population in $\left|a\right>$, leading, from equation~(\ref{rhotf}), to:
\begin{equation} \label{eqdepa}
p_a=\sum_n p_n p_n^a = \frac{1}{2}\left\{ 1 - C(t_f)\cos{\left[\Delta\varphi(t_f)\right]} \right\}\;,
\end{equation}
where we introduce the contrast:
\begin{equation}
C(t) = \left|A(t)\right| \;, 
\label{ContrastGene}
\end{equation}
and the phase-shift: 
\begin{equation}
\Delta\varphi(t) = \arg{\left[ A(t) \right]} \;,
\label{PhaseShiftGene}
\end{equation}
with $A(t) = \sum_n p_n \exp(j\delta\phi - j\omega_{ab}t -  j(\Omega_n^b-\Omega_n^a))$.

\section{Phase-shift and contrast in the adiabatic case}
\label{sec_adiabatic}

In this section, we assume that the time variations of $V_a(\widehat{z},t)$ and $V_b(\widehat{z},t)$ are slow enough that the adiabatic approximation can be applied, as discussed in \cite{Ammar2014}. A more general non-adiabatic case will be considered in section \ref{sec_beyond_adiabatic}. We furthermore assume that the path in parameter space describing the changes in $V_{a,b}(\widehat{z},t)$ retraces itself, such that the geometrical phase factors vanish \cite{Berry1984} and thus $\Omega_n^i = \int_0^{t_f} E_n^i(t) dt / \hbar$ where $E_n^i(t)$ are the adiabatic eigen-energies of $\widehat{H} \left|i \right>\left<i\right|(t)$. Moreover, we assume for simplicity that the duration of the splitting and merging period are much smaller than the duration of the interrogation period, such that the effect of splitting and merging on the phase shift and contrast can be neglected (taking into account more realistic interferometric sequences, as described in \cite{Ammar2014}, does not change the conclusions drawn in this section).
We can thus write the phase accumulated by $\left|n_i(t)\right>\left|i\right>$ as $\Omega_n^i = E_n^i t_f /\hbar$ leading to $A(t) = \sum_n p_n \exp(j\omega t_f - j\omega_{ab}t_f -  j\delta\omega_n t_f)$ where $\delta\omega_n = (E_n^b - E_n^a)/\hbar$ is difference between the eigen energies of the two traps for the same vibrational level.

\subsection{Rule of thumb for the coherence time}
\label{subsec_coherence_time}

A very convenient rule of thumb to infer the coherence time can be derived from equation~(\ref{ContrastGene}) by considering the second order Taylor expansion of $C$ under the assumption $\left| \delta \omega_n \right| t \ll 1$. This leads to $C(t) \simeq 1-\left(t/t_c\right)^2/2$, where $t_c$ is understood as the coherence time, with the following expression for $t_c$:
\begin{equation}
t_c \simeq \left[ \sum_n p_n \delta \omega_n^2 - \left( \sum_n p_n \delta \omega_n \right)^2 \right]^{-1/2}\;.
\label{CoherenceTime}
\end{equation}
In other words, the inferred decoherence rate $t_c^{-1}$ is on the same order of magnitude as the standard deviation of the $\delta \omega_n$, weighted by the Boltzmann factors $p_n$.

If we furthermore assume that $V_a$ and $V_b$ correspond, during the interrogation period, to two harmonic trap with slightly different frequencies $\omega_a$ and $\omega_b$, with $\left|\omega_a-\omega_b\right| \ll \omega_{a,b}$, equation (\ref{CoherenceTime}) leads, in the case of a weakly degenerate gas $\hbar \omega_{a,b} \ll kT$, to:
\begin{equation} \label{tcstaticharmonique} 
t_c \simeq \frac{1}{\delta\omega} \frac{\hbar\omega}{kT} \;,
\label{tcdw}
\end{equation}
with $\omega = \left(\omega_a+\omega_b\right)/2$ and $\delta\omega = |\omega_a-\omega_b|$. It is obvious from equation~(\ref{tcdw}) that $t_c$ increases with symmetry and decreases with temperature, as expected intuitively. This result differs from the exact calculation, in case of two harmonic potentials \cite{Ammar2014}, only by a factor $\sqrt{3}$. For a typical temperature of 500~nK, equation~(\ref{tcdw}) gives a symmetry-limited coherence time on the order of 15~ms for a realistic value of the asymmetry $\delta\omega/\omega \lesssim $~10$^{-3}$ \cite{DupontNivet2016}. In the case of non-harmonic traps, equations (\ref{ContrastGene}) or (\ref{CoherenceTime}) can still be used with perturbatively - or numerically - estimated values of the eigen-energies.

\subsection{Phase-shift in the presence of a gravity or acceleration field}
\label{subsec_phase_shift}

In the rest of this paper, we consider the case where $V_i\left(\widehat{z}\right)$ is the sum of a harmonic potential and an acceleration or gravity potential namely:
\begin{eqnarray}
V_i\left(\widehat{z}\right) & = & \frac{m\omega_i^2}{2}\left( \widehat{z} -z_i\right)^2 + mg\widehat{z} 	  \nonumber \\
& = & \frac{m\omega_i^2}{2}\left( \widehat{z}-z_i^{cm} \right)^2 + \frac{mg^2}{2\omega_i^2} + mgz_i^{cm}
\label{HStatic}
\end{eqnarray}
where $m$ is the atomic mass, $\omega_i$ are the trap frequencies, $g$ is the acceleration or gravity field, $z_i$ is the trap center (minimum of the trapping part of the potential) and $z_i^{cm}=z_i - g/\omega_i^2$ is the center of mass position of the atoms. The phase difference $\Delta\varphi(t)$ (equation (\ref{PhaseShiftGene})) after an interrogation time $t$, stemming from Hamiltonian (\ref{hamiltonian}) and potential (\ref{HStatic}), is given in this case by:
\begin{equation}
\Delta\varphi(t) = (\omega-\omega_{ab})t - \Delta\varphi^0(t) \;,
\label{eq_PhaseShift} 
\end{equation}
with:
\begin{eqnarray}
\Delta \varphi^0 (t) & = & \left[ mg\left(z_b^{cm}-z_a^{cm} \right) + \frac{mg^2}{2} \left( \frac{1}{\omega_b^2} - \frac{1}{\omega_a^2} \right) \right]  \frac{t}{\hbar}    \nonumber   \\
 & + & \frac{\omega_b - \omega_a}{2} t + \phi_T(t) 
\label{SignalStatic}
\end{eqnarray}
where : 
\begin{equation}
\phi_T(t) = \arctan \left\{ \frac{\sin\left( \left(\omega_b-\omega_a\right)t \right)e^{-\hbar\omega_a/(kT)}}{1-\cos\left( \left(\omega_b-\omega_a\right)t \right)e^{-\hbar\omega_a/(kT)}} \right\} \;.
\label{PhaseStaticTemperature}
\end{equation}
In equation (\ref{eq_PhaseShift}) $\Delta\varphi^0(t)$ arises from the spatial separation of the two internal states, and $(\omega-\omega_{ab})t$ describe the free evolution of the states. In equation (\ref{SignalStatic}), the first term is the classical difference in potential energy due to the presence of the acceleration or gravity field. The second term is an energy shift resulting from the addition of the harmonic potential with the linear $ mg\widehat{z}$ term (see equation (\ref{HStatic})). The third term is the difference of zero point energies of the two harmonic oscillators. The last term, which is temperature dependent, vanishes in two cases : i) a symmetric interferometer (i.e. $\omega_a=\omega_b$), ii) zero-temperature. Equation (\ref{PhaseStaticTemperature}) shows that not only the contrast depends on temperature (as was predicted in \cite{Ammar2014}) but also the phase-shift. We also predict a direct link between the phase-shift and the relative asymmetry of the two traps, as was previously pointed out in \cite{Sidorov2006}.

\section{Beyond the adiabatic case : shortcuts to adiabadicity (STA)}
\label{sec_beyond_adiabatic}

Let us now consider the dynamical problem of splitting and recombination. As illustrated by the numbers given previously for the coherence time, it is not always possible to perform adiabatic splitting and recombination (which have to be longer than the trap period \cite{Ammar2014}), because the inverse of the inferred coherence time ($\simeq 15$~ms) is on the same order of magnitude as usual trapping frequencies in atom chip experiments (typically between 10~Hz and 1~kHz \cite{reichel2010}). 

\subsection{Shortcut to adiabadicity ramps}
\label{subsec_STAramps}

It has been demonstrated in \cite{Torrontegui2011,Schaff2011} that non-trivial temporal ramps can be used to move an atomic cloud while keeping the population of the different quantum levels unchanged at the ends of the ramp, on the time scale of the trapping period (hence much faster than an adiabatic ramp \cite{Ammar2014}). We propose, in the following, to apply this technique, known as shortcut to adiabadicity \cite{Torrontegui2013,Schaff2011,Chen2010,Torrontegui2011} (STA), to the case of a trapped thermal atom interferometer. For simplicity, we only consider the case of a harmonic trap (for other potentials the reader is referred to \cite{Torrontegui2013} and references therein). We thus consider a trapping potential with a time-depend position and stiffness:
\begin{equation}
V_{i}\left(\widehat{z},t\right) = \frac{m\omega_{i}^2(t)}{2} \left[ \widehat{z} - z_{i}(t) \right]^2 + mg\widehat{z} \;.
\end{equation}
Similar to the case of equation (\ref{HStatic}), we can rewrite these potentials as:
\begin{eqnarray}
V_{i}\left(\widehat{z},t\right) & = &  \frac{m\omega_{i}^2(t)}{2}\left[ \widehat{z} - z_{i}(t)  + \frac{g}{\omega_{i}^2(t)} \right]^2 + \gamma_{i}(t)  \nonumber\\
\mathrm{with :}   & & \gamma_{i}(t)  =  - \frac{mg^2}{2\omega_{i}^2(t)} + mgz_{i}(t) \;.
\end{eqnarray}
To introduce the mathematical condition which must be fulfilled for the STA, we need to write a dynamical invariant $\widehat{I}_{i}(t)$ of $\widehat{H}\left|i\right>\left<i\right|(t)$. $\widehat{K}$ is a dynamical invariant of an operator $\widehat{P}$ if \cite{Lewis1969}: i) $j\hbar\partial_t \widehat{K} + [\widehat{K}, \widehat{P} ]=0$ and ii) $\widehat{K}$ is hermitian. Expressions for $\widehat{I}_i(t)$ can be found in the literature \cite{Lewis1982,Dhara1984,Schaff2011}. After adapting them to include the presence of $g$, we obtain:
\begin{eqnarray}
\widehat{I}_i(t) & = & \frac{\omega_0}{2m} \left[ \rho_i\left( \widehat{p}-m\dot{z}_i^{cm} \right) - m\dot{\rho}_i\left( \widehat{z} - z_i^{cm}\right) \right]^2 + \frac{m\omega_0}{2}\frac{\left( \widehat{z}-z_i^{cm}\right)^2}{\rho_i^2}	 \;,  \label{invariant}
\end{eqnarray}
where $\omega_0$ is an arbitrary angular frequency and $\rho_i$ and $z_i^{cm}$ are solutions of the following equations:
\begin{eqnarray}
& & \ddot{\rho}_i + \omega_i^2(t)\rho_i = \frac{1}{\rho_i^3} \;,	 \label{eqrho} \\
& & \ddot{z}_i^{cm} + \omega_i^2(t) \left[ z_i^{cm} - z_i(t) + \frac{g}{\omega_i^2(t)}\right] = 0 \;.	\label{eqzi}
\end{eqnarray}
Equation (\ref{eqrho}) is the Ermakov equation and equation (\ref{eqzi}) is the classical linear oscillator. Physically, $z_i^{cm}$ is the center of mass of the atomic cloud obeying equation~(\ref{eqzi}), and $\rho_i$ is proportional to the cloud size \cite{Schaff2011}. For a given time $t=t_p$, the populations of the different quantum levels will be the same as for $t=t_m$ if $\widehat{H}\left|i\right>\left<i\right|(t_m)\propto \widehat{I_i}(t_m)$ and $\widehat{H}\left|i\right>\left<i\right|(t_p)\propto \widehat{I_i}(t_p)$~\cite{Schaff2011,Torrontegui2013}. This imposes in particular the following conditions on $\rho_i$ and $z_i^{cm}$ at $t_{m,p}$:
\begin{eqnarray}
\rho_i(t_{m,p}) = \frac{1}{\sqrt{\omega_i(t_{m,p})}}\;,  	& \qquad  & z_i^{cm}(t_{m,p}) = z_i(t_{m,p}) - \frac{g}{\omega_i^2(t_{m,p})} \;,  \nonumber \\
\dot{\rho}_i(t_{m,p}) = 0 \;,						& \qquad  & \dot{z}_i^{cm}(t_{m,p}) = 0 \;,  \label{STAcondition}
\end{eqnarray}
where $\omega_i(t_{m,p})$ and $z_i(t_{m,p})$ are fixed parameters which are linked to the equilibrium position and cloud size at $t_{m,p}$.  Two additional conditions: $\ddot{\rho}_i(t_{m,p}) = 0$ and $\ddot{z}^{cm}_i(t_{m,p})=0$ are provided by (\ref{eqrho}) and (\ref{eqzi}). Together with (\ref{STAcondition}) they form the STA conditions at $t_{m,p}$.

\begin{figure}
\centering  \includegraphics[width=0.7\textwidth]{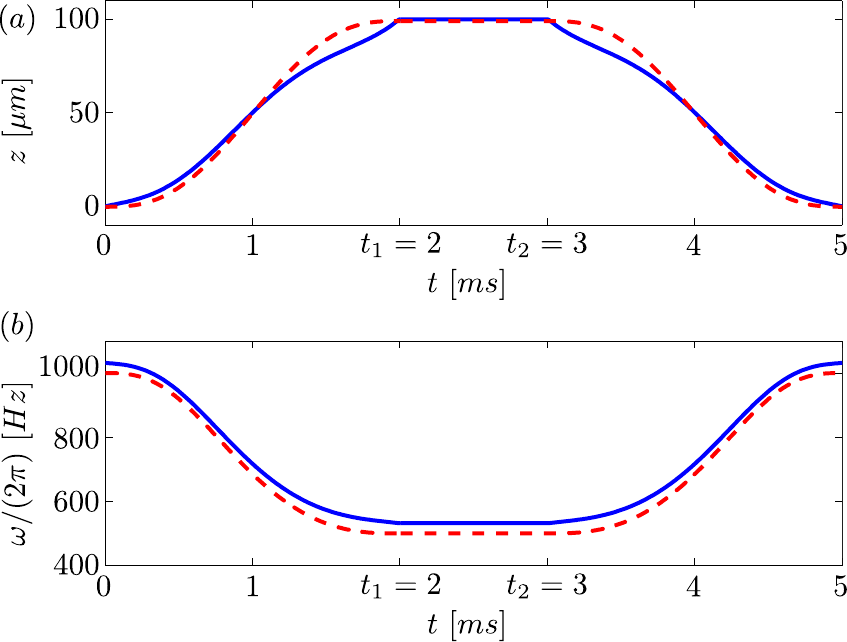}
\caption{\label{FigTimeRamp} a) Representation of temporal ramps for $z_i$ (blue solid line), $z_i^{cm}$ (red dashed line). b) Representation of the temporal ramp for $\omega_i$ (blue solid line) and $1/\rho_i^2$ (red dashed line). The (a) blue solid line, corresponding to $\omega_i$, has been translated 30~Hz upwards for readability. Between $t=0$ and $t=t_1$, we spatially separate by 200 $\mu$m the two internal states $\left|a\right>$ and $\left|b\right>$ (splitting period). During this phase the trap frequency $\omega_i$ is decreased from 1~kHz to 500~Hz. In our numerical example $t_1=$~2~ms, and we require STA conditions at $t=0$ and $t=t_1$. Between $t_1$ and $t_2$ (interrogation period), the frequencies and trap positions are held constant. Between $t_2$ and $t_f$ we spatially recombine the two states (merging period). For simplicity, we show the motion of one well only. For the other one, the frequency ramp is the same and the spatial motion is assumed to be in the opposite direction.} 
\end{figure}

In order to find a temporal ramp on $\omega_i$ and $z_i$ for the splitting, we need to solve equations (\ref{eqrho}), (\ref{eqzi}) and (\ref{STAcondition}). To do this, as we have six conditions on $\rho_{i}$ and six on $z_i^{cm}$, we take a fifth-order polynomial ansatz for $\rho_i$ and $z_i^{cm}$ \cite{Schaff2011,Chen2010,Torrontegui2011}. The frequency ramp is first found from $\rho_i$ and (\ref{eqrho}) and the trap position $z_i$ is then deduced from $\omega_i$, $z_i^{cm}$ and (\ref{eqzi}). To give a numerical example, the following parameters are taken (times are defined in figure \ref{FigTimeRamp}): $t_1=$~2~ms, $\omega_i(0)/2\pi=$~1~kHz, $\omega_i(t_1)/2\pi=$~500~Hz, $g$ is the gravitational acceleration and the maximum separation distance between the two internal states is 200~$\mu$m. This example is shown in figure \ref{FigTimeRamp}, where we use the same ramp for recombination and splitting. Numerically we were not able to find $t_1$ significantly lower than 2~ms while preserving a smooth ramp for the frequency (without imaginary frequencies to keep the trapping behaviour of the potential) and for the trap position. This is in accordance with \cite{Salamon2009} where it is stated that the minimum time is on the order of $2\pi/\omega_i$.

\subsection{Contrast and phase-shift with STA ramps}
\label{subsec_contrast_phase_shift}

For purposes of interferometry, the contribution to the overall phase shift of the splitting and merging period has to be taken into account, all the more since their duration is not negligible compared to the typical value of the coherence time inferred previously. The framework of the dynamical invariant $\widehat{I_i}(t)$ \cite{Lewis1969} provides a tool to compute this overall phase shift between $t=0$ and $t=t_f$ (i.e. during the whole interferometric sequence). Reference \cite{Lewis1969} gives the following generic solution $\left|t\right>\left|i\right>$ of the Schr\"odinger equation with a time-dependent hamiltonian $\widehat{H}\left|i\right>\left<i\right|(t)$:
\begin{eqnarray}
\left|t\right>\left|i\right> & = & \sum_n c_n^i \exp{ \left( j\alpha_n^i(t) \right) } \left| n(t)\right>\left|i\right> \;,
\end{eqnarray}
where $c_n^i$ are time-independent factors which depend on the initial conditions, $\left|n(t)\right>\left|i\right>$ are the eigen-states of $\widehat{I}_i(t)$ and the $\alpha_n^i(t)$ are chosen such that $\exp{ \left( j\alpha_n^i(t) \right) } \left|n(t)\right>\left|i\right>$ are solutions of the Schr\"odinger equation with the hamiltonian $\widehat{H}\left|i\right>\left<i\right|(t)$ \cite{Lewis1969}. Adapting the results of \cite{Schaff2011,Popov1969,Popov1970} to the case of the trapped interferometer considered in this paper, we obtain:
\begin{eqnarray}
\arg{\left( \exp{ \left( j\alpha_n^i(t) \right) } \left|n(t)\right>\left|i\right>\right)} = & - & \left(n+\frac{1}{2}\right)\int_0^t \frac{dt'}{\rho^2_i(t')} + \Psi_i(z,t)  \nonumber\\
&  - &  \frac{F_i(t)}{\hbar} - \frac{\Gamma_i(t)}{\hbar} \;,
\label{SolGeneTimeDependent}
\end{eqnarray}
with the following expressions for $\Psi_i$, $F_i$ and $\Gamma_i$:
\begin{eqnarray}
\Psi_i(z,t) 		& = & \frac{m}{\hbar}\left[ \frac{\dot{\rho_i}}{2\rho_i}z^2 + \frac{1}{\rho_i}\left( \dot{z}_i^{cm}\rho_i - z_i^{cm}\dot{\rho} \right)z \right]  	\nonumber\\
F_i(t) 			&  = & \frac{m}{2}\int_0^t dt' \left[  \frac{1}{\rho_i^2}\left( \dot{z}_i^{cm}\rho_i - z_i^{cm}\dot{\rho} \right)^2 \right]  						\nonumber\\
		 	& + &  \frac{m}{2}\int_0^t dt' \left[ - \frac{\left( z_i^{cm} \right)^2}{\rho_i^4}  + \omega_i^2 \left( z_i - \frac{g}{\omega_i^2} \right)^2 \right]			\nonumber \\
\Gamma_i(t)		& = &  \frac{m}{2} \int_0^t dt' \left[ 2gz_i - \frac{g^2}{\omega_i^2} \right] \;.
\label{PhaseSolGene}
\end{eqnarray}

As the $\exp{ \left( j\alpha_n^i(t) \right) } \left|n(t)\right>\left|i\right>$ are all solutions of the Schr\"odinger equation with the Hamiltonian $\widehat{H}\left|i\right>\left<i\right|(t)$ and form a orthonormal basis of our Hilbert space \cite{Lewis1969,Schaff2011,Popov1969,Popov1970}, we can easily extend equation (\ref{rhotf}) to account for time-dependent splitting and recombination. Thus from (\ref{SolGeneTimeDependent}) and (\ref{PhaseSolGene}) we can compute the contrast and the phase-shift. In this case the term $\Omega_n^b-\Omega_n^a$ from the definition of $A(t)$ is equal to: $\arg(\exp(j\alpha_n^b(t))\left|n(t)\right>\left|b\right>)-\arg(\exp(j\alpha_n^a(t))\left|n(t)\right>\left|a\right>)$. Under the same hypothesis as in the adiabatic case (equation (\ref{tcstaticharmonique})), the coherence time $t_c$ can be inferred by solving the following equation:
\begin{equation}
\sqrt{3} = \frac{kT}{\hbar\omega} \left| \int_0^{t_c} \left( \frac{1}{\rho_a^2} - \frac{1}{\rho_b^2} \right) dt \right| \;,
\end{equation}
which is a dynamical version of equation (\ref{tcdw}) for time-dependent frequencies. It is interesting to notice that $z_i(t)$ has no role in this expression, which is consistent with the fact that a translation or a rotation of an Hamiltonian preserves its eigen-values, and thus it preserves the contrast as already pointed out in \cite{Ammar2014}.

As regards the phase-shift $\Delta\varphi(t)$, only the splitting dependent part $\Delta\varphi^0(t)$ changes and it is given by: $\Delta\varphi^0(t) = \Psi_a(z,t) - \Psi_b(z,t) - F_a(t)/\hbar + F_b(t)/\hbar - \Gamma_a(t)/\hbar+\Gamma_b(t)/\hbar - \frac{1}{2}f(t) + \arg\left[ \sum_n p_n \exp\left( -jnf(t) \right) \right]$, with $f(t)=\int_0^t 1/\rho_a^2 dt'-\int_0^t 1/\rho_b^2 dt'$. Assuming that STA conditions are fulfilled at $t=0$ and $t=t_f$ \footnote{Only the conditions $\dot{\rho}_{a,b}(0)=\dot{\rho}_{a,b}(t_f)=0$ and $\dot{z}_{a,b}^{cm}(0)=\dot{z}_{a,b}^{cm}(t_f)=0$ are needed.}, we obtain the following (more explicit) expression for the phase-shift : $\Delta\varphi(t_f)= (\omega-\omega_{ab})t_f - \Delta\varphi^0(t_f)$, with:
\begin{eqnarray}
\Delta\varphi^0(t_f) & = & \frac{m}{2\hbar} \int_0^{t_f} \left[ \left( \dot{z}_a^{cm} \right)^2  -  \left( \dot{z}_b^{cm} \right)^2 \right] dt  		\nonumber\\
& - &    \frac{mg}{\hbar} \int_0^{t_f}  \left( z_a^{cm} - z_b^{cm} \right) dt		 \nonumber\\
& - & \frac{m}{2\hbar} \int_0^{t_f} \left[ \left( \frac{\ddot{z}_a^{cm}+g}{\omega_a} \right)^2 -  \left( \frac{\ddot{z}_b^{cm}+g}{\omega_b} \right)^2 \right] dt  	\nonumber\\
&  - & \frac{1}{2} \int_0^{t_f} \left( \frac{1}{\rho_a^2} - \frac{1}{\rho_b^2} \right) dt  - \phi_T(t_f)
\label{DynamicPhaseShift}
\end{eqnarray}
where : 
\begin{equation}
\phi_T(t_f) = \arctan \left\{ \frac{\sin\left( f(t_f) \right)e^{-\hbar\omega_a/(kT)}}{1-\cos\left( f(t_f) \right)e^{-\hbar\omega_a/(kT)}} \right\} \;.
\end{equation}

In equation (\ref{DynamicPhaseShift}), the first term comes from kinetic energy. The second is the classical difference in potential gravitational energy. The third comes from the energy shift of the harmonic oscillator levels in the presence of the overall acceleration field of the atomic cloud $g+\ddot{z}_i^{cm}$ (i.e. acceleration of the whole interferometer and acceleration of the trap). The fourth term comes from the difference in zero point energies of the two oscillators. To make the latter more explicit, we point out that in the case where $\omega_i$ is time-independent, then $1/\rho_i^2=\omega_i$ and the fourth term of equation (\ref{DynamicPhaseShift}) becomes identical to the third term of equation~(\ref{SignalStatic}). The last term includes the temperature dependence of the phase shift and it is the analogue of (\ref{PhaseStaticTemperature}) for the time dependent case.

\subsection{Towards an accelerometer ?}
\label{subsec_accelerometer}

In a practical implementation of this interferometer \cite{Ammar2014}, the experimental parameters are $\omega_i$ and $z_i$, and not $\rho_i$ and $z_i^{cm}$. From the two STA ramps for $\rho_i$ and $z_i^{cm}$, we need to compute the ramps for the two experimental parameters $\omega_i$ and $z_i$. In the general case, the computation of $z_i$ requires the knowledge of $g$, which is the parameter we want to measure. This circle can be broken in the two following cases : 

i) We choose the splitting time $t_1$ and the trap frequency $\omega_i$ such that $z_i^{cm} \simeq z_i$. If we call $d$ the splitting distance, the latter choice and equation (\ref{eqzi}) imply that $t_1^2\omega_i^2 \gg 1$ and $g/(\omega_i^2d) \ll 1$, i.e. an adiabatic splitting and a strong trap confinement to make the acceleration shift of the trap position negligible. In this ideal adiabatic case, the phase-shift $\Delta \varphi^0 (t_f)$ reduces to:
\begin{equation}
\Delta\varphi^0(t_f) = - \frac{mg}{\hbar} \int_0^{t_f}  \left( z_a - z_b \right) dt	 \;,
\label{PhaseShiftSimple}
\end{equation}
making such a system an attractive candidate for acceleration measurements. Assuming a phase measurement limited by the quantum projection noise leads to an uncertainty on the measurement of $g$ on the order of $\delta g/g \sim \hbar/ m\Delta z t_c \sqrt{N}$ per shot. For example, with the following numerical values: $\Delta z \sim$~100~$\mu$m, $t_c\sim$~10~ms, $N\sim$~1000 atoms and $m=1.4$~$10^{-25}$~kg for ${}^{87}$Rb we obtain  $\delta g/g=$2$\cdot$10${}^{-6}$ per shot.

ii) If the adiabatic approximation is not valid for example because of a too short coherence time, it is still possible to use the previously described interferometer to measure an acceleration. In the case of identical time-dependent-stiffness for the two traps, i.e. $\rho_a=\rho_b$, we suppose that a time-dependent function $h$ exists and satisfies the two following conditions: 1)~$z_a^{cm}=(d-g/\omega_s^2+g/\omega_r^2)h-g/\omega_r^2$ and $z_a^{cm}=(-d-g/\omega_s^2+g/\omega_r^2)h-g/\omega_r^2$ where $\omega_r=\omega(0)=\omega(t_f)$, $\omega_s=\omega(t_1)=\omega(t_2)$ and $d=|z_a(t_1,t_2)|=|z_b(t_1,t_2)|$ and 2)~the STA conditions are fulfil for $z_a^{cm}$ and $z_b^{cm}$. The important point is that finding such a function $h$ does not imply the knowledge of the acceleration $g$. In this case, the time dependent-splitting distance is $z_a-z_b=2d\ddot{h}/\omega^2+2dh$ and this last function can be used in the interferometer sequence to measure the acceleration $g$.

\section{Conclusion}

To summarize, we have given in this paper some quantitative elements to estimate the required degree of symmetry to implement an interferometer with trapped thermal atoms, and the associated phase shift taking into account the acceleration field $g$ and the splitting dynamics. The inferred coherence time roughly scales with the inverse of the variance of the energy difference of the levels of the two traps, weighted by the Boltzmann distribution. Taking the example of two harmonic traps, we find that a coherence time of $15$~ms could be achieved if the symmetry is controlled to better than $10^{-3}$. Remarkably in the presence of a dynamic splitting the contrast retain approximatively the same form. We also derived expression for the phase shift and contrast in the dynamical case based on the STA formalism, showing that splitting and recombination could be achieved on time scale of the same order of magnitude as the trapping period.

One promising way to achieve the high degree of symmetry inferred in this paper is on-chip Ramsey interferometry with the clock states of the ${}^{87}$Rb, as described in references \cite{Bohi2009,Ammar2014}, because it provides a quasi-independent control on the potentials of the internal states, especially if two coplanar wave guides are used to address independently the two internal states \cite{Ammar2014}. This formalism could also be applied to interferometers using cold fermions \cite{Roati2004}, in which case atom interaction effects are negligible.

\ack
This work has been carried out within the OnACIS project ANR-13-ASTR-0031 funded by the French National Research Agency (ANR) in the frame of its 2013 Astrid program.

\section*{References}
\bibliographystyle{unsrt}
\bibliography{biblio_these}

\end{document}